\documentclass[12pt]{article}
\usepackage{helvet}
\usepackage{color}
\usepackage{fancyhdr}
\usepackage{float}
\usepackage{vmargin}
\usepackage{amsmath,amsfonts,amssymb}
\usepackage{mathtools}
\usepackage{graphicx}
\usepackage{lastpage}
\usepackage{here}
\usepackage{tabularx}
\usepackage{tabulary}
\usepackage{booktabs}
\usepackage{hyperref}
\usepackage{algorithm}
\usepackage[font=footnotesize]{caption}
\usepackage{subcaption}
\usepackage[noend]{algpseudocode}
\usepackage[table]{xcolor}
\usepackage{graphicx}
\usepackage[utf8]{inputenc}
\usepackage{setspace}
\usepackage{wrapfig}
\usepackage[comma,numbers]{natbib}

\usepackage{lipsum}

\newcommand{\tte}[1]{\textcolor{red}{\Comment{\\tte: #1\\}}}
\newcommand{\lth}[1]{\textcolor{blue}{\Comment{\\lth: #1\\}}}
\newcommand{\dow}[1]{\textcolor{green}{\Comment{\\dow: #1\\}}}
\newcommand{\mrm}[1]{\textcolor{purple}{\Comment{\\mrm: #1\\}}}

\newcommand{\beginsupplement}{%
        \setcounter{section}{1}
        \setcounter{subsection}{0}
        \renewcommand{\thesubsection}{S.\arabic{subsection}}%
        \setcounter{page}{1}
        \setcounter{table}{0}
        \renewcommand{\thetable}{S\arabic{table}}%
        \setcounter{figure}{0}
        \renewcommand{\thefigure}{S\arabic{figure}}%
        \setcounter{algorithm}{0}
        \renewcommand{\thealgorithm}{S\arabic{algorithm}}%
     }

\begin{document}
 
\begin{Large}
\noindent \textbf{PIMKL: Pathway Induced Multiple Kernel Learning}
\end{Large}\\
\begin{footnotesize}
\noindent Matteo Manica\textsuperscript{1,2,*}, Joris Cadow \textsuperscript{1,2,*}, Roland Mathis \textsuperscript{1,*}, María Rodríguez Martínez\textsuperscript{1, $\dagger$}\\
\noindent \textit{ \{tte, dow, lth, mrm\}@zurich.ibm.com}\\
\textsuperscript{1} IBM Research Zürich\\
\textsuperscript{2}	ETH - Zürich\\ 
\textsuperscript{*} Shared first authorship \\
\textsuperscript{ $\dagger$} Corresponding author
\end{footnotesize}
\section*{Abstract}

\subsection*{Motivation}
Reliable identification of molecular biomarkers is essential for accurate  patient stratification. While state-of-the-art machine learning approaches for sample classification continue to push boundaries in terms of performance, most of these methods are not able to integrate different data types and lack generalization power, limiting their application in a clinical setting.
Furthermore, many methods behave as black boxes, and we have very little understanding about the mechanisms that lead to the prediction. While opaqueness concerning machine behaviour might not be a problem in deterministic domains, in health care, providing explanations about the molecular factors and phenotypes that are driving the classification is crucial to build trust in the performance of the predictive system. 

\subsection*{Results}
We propose Pathway Induced Multiple Kernel Learning (PIMKL), a novel methodology to reliably classify samples  that can also help gain insights into  the molecular mechanisms that underlie the classification. PIMKL exploits prior knowledge in the form of a molecular interaction network and annotated gene sets, by optimizing a mixture of \textit{pathway-induced} kernels using a Multiple Kernel Learning (MKL) algorithm, an approach that has demonstrated excellent performance in different machine learning applications. After optimizing the combination of kernels for prediction of a specific phenotype, the model provides a stable molecular signature that can be interpreted in the light of the ingested prior knowledge and that can be used in transfer learning tasks.


\subsection*{Contact}
mrm@zurich.ibm.com

\vspace{1\baselineskip}

\noindent \textbf{Keywords:} Biological  Networks, Pathways, Machine Learning,  Patient Stratification,  Multiple Kernel Learning.
\section{Introduction}
\label{sec:introduction}

Designing reliable and interpretable predictive models for patient stratification and biomarker discovery is a daunting challenge in computational biology. A plethora of methods based on molecular data have been proposed throughout the years, many of which exploit prior knowledge about the molecular processes involved in the regulation of the phenotype to be predicted. Prior knowledge  is frequently encoded as a molecular interaction network, where nodes represent genes or proteins and edges represent  relationships between the connected nodes.
Supporting the development of such  methods, the number of databases reporting protein-protein interactions has seen an unprecedented growth in recent years, and databases such as STRING \cite{Szklarczyk2017}, OmniPath \cite{Turei2016}, Reactome \cite{Croft2014,Fabregat2018}, IntAct \cite{Kerrien2011}, MINT \cite{Licata2011}, MatrixDB \cite{Chautard2009}, HPRD \cite{Keshava2008}, KEGG \cite{Zhang2009,Tenenbaum2017,Kanehisa2000} or Pathway Commons \cite{Cerami2011}, just to name a few, provide an incredibly useful resource for designing models informed about the underlying molecular processes.

Several studies have focused on comparing prior knowledge-based classification methods. For instance, Cun and Fr\"ohlich~\cite{Cun2012} evaluated 14 machine learning approaches to predict the  survival outcome of breast cancer patients. The methods included, among others: average pathway expression \cite{Guo2005}, classification by significant hub genes \cite{Taylor2009}, pathway activity classification \cite{Lee2008}, and a series of approaches based on Support Vector Machines (SVMs), such as network-based SVMs \cite{Zhu2009}, recursive feature elimination SVMs \cite{Guyon2002}, and graph diffusion kernels for SVMs \cite{Rapaport2007,Gao2009}.
The study concluded that, while none of the evaluated approaches  significantly improved classification accuracy, the interpretability of the gene signatures obtained was greatly enhanced by the integration of prior knowledge.

A more recent benchmarking effort was provided by a collaboration between the National Cancer Institute (NCI) and the Dialogue on Reverse Engineering Assessment and Methods (DREAM) project \cite{Costello2014}. The NCI-DREAM challenge aimed to identify the top-performing methods for predicting therapeutic response in breast cancer cell lines using genomic, proteomic, and epigenomic data profiles. A total of 44 prediction algorithms were scored against an unpublished and hidden gold-standard data set. Two interesting conclusions emerged from the challenge. First, all top-performing methods modeled nonlinear relationships and incorporated biological pathway information, and second, performance was increased by including multiple, independent data sets. Interestingly,  the top-performing methodology, Bayesian Multitask Multiple Kernel Learning, exploited a multiple kernel learning (MKL) framework \cite{Gonen2011}.

MKL methods aim to model complex and heterogeneous datasets by using a weighted combination of base kernels. While in more traditional kernel methods the parameters of a single kernel are optimized during training, in MKL, the weights of each kernel are tuned during training. 
Compared to single-kernel methods, the advantages of MKL are two-fold. First, different kernels can encode various levels of information, e.g. different definitions of similarity or different types of data, endowing the algorithm with the flexibility required to model heterogeneous or multi-modal datasets. Second, after optimizing the combination of kernels, the weights associated with each kernel can provide valuable insights about the sets of features that are most informative for the classification task at hand.

In this paper, we seek to augment the predictive power and interpretability of MKL methods, by supplementing them with the use of prior knowledge. Towards this end, we introduce the Pathway Induced Multiple Kernel Learning (PIMKL), a supervised classification algorithm for phenotype prediction from molecular data that jointly exploits the benefits of MKL and prior knowledge ingestion. PIMKL uses an interaction network and a set of annotated gene sets to build a mixture of \textit{pathway-induced} kernels from molecular data, whose mixture is then  optimized with an MKL algorithm.
After PIMKL is trained, the weight assigned to each kernel provides information about the importance of the corresponding pathway in the mixture. As a result, a molecular signature characterizing the phenotype of interest is derived.

While there are currently many approaches that take advantage of the known graph structure of a molecular system~\cite{jacob2012,Rapaport2007}, or use collections of annotated gene sets as prior knowledge to
reduce the dimensionality of molecular profiles and enable the analysis of tumour profiles~\cite{livshits_pathwaybased_2015,chang_pathwaybased_2015}, to our knowledge PIMKL is the first methodology that integrates both levels of prior knowledge -- molecular networks and collections of pathways -- with state-of-the-art machine learning approaches. We demonstrate that the use of MKL enhances classification performance, and the use of prior knowledge ensures that the results are interpretable and shed light on the molecular interactions implicated in the phenotype. 

This paper is structured as follows. We first describe PIMKL and validate it by predicting disease-free survival for breast cancer samples from multiple cohorts. We benchmark PIMKL by comparing it with the methods analyzed in \cite{Cun2012}. To evaluate its generalization power, we use a PIMKL-generated molecular signature to predict disease-free survival on a different dataset, the METABRIC breast cancer cohort \cite{Curtis2012}. 
Finally, we test PIMKL  robustness against noise and its capabilities to integrate distinct data types by simultaneously using  METABRIC gene expression (mRNA) and copy number alteration (CNA) data for the same classification task.
Our analysis suggest that PIMKL provides an extremely robust approach for the integration of multiple types of data with prior knowledge that can be successfully applied to a wide range of phenotype prediction problems.

\section{Methods}
\label{sec:methods}

PIMKL is a methodology for phenotype prediction from multi-omic measurements, e.g. mRNA, CNA, etc, based on the optimization of a mixture of \textit{pathway-induced} kernels. Such kernels are generated by exploiting prior knowledge in a dual fashion. First, prior knowledge is injected in PIMKL in the form of a molecular interaction network, and second, as a set of annotated gene sets or pathways.

\begin{figure}[!htb]
\centering
\includegraphics[width=0.85\textwidth, clip]{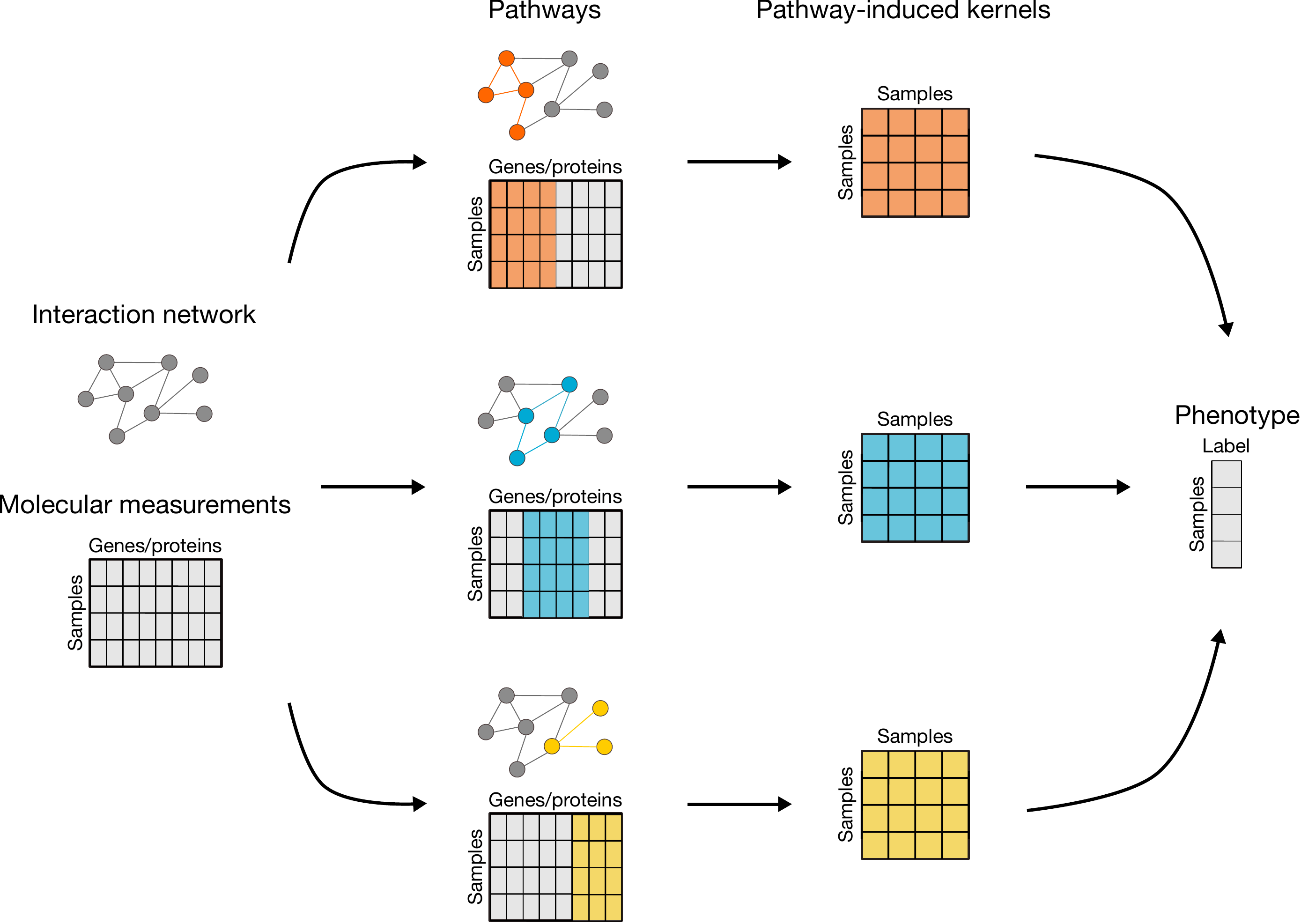}\\
\caption{
\textbf{PIMKL concept.}
Given a network topology describing molecular interactions, relevant sub-networks can be extracted to generate a mixture of \textit{pathway-induced} kernels. The combination of kernels is then optimized to predict a phenotype of interest. The weights of the mixture provide a measurement of the importance of each  pathway, thereby shedding light on the molecular mechanisms that contribute to the  phenotype.
}
\label{fig:pimkl_concept}
\end{figure}

A key aspect of PIMKL is \textit{pathway induction}, a method to generate similarity functions using the topological properties of an interaction network. In practice, we use pathway gene sets with well-defined biological functions to define sub-networks from which we generate \textit{pathway-induced} kernels. This mixture of \textit{pathway-induced} kernels is then optimized to classify a phenotype of interest, and in doing so, each pathway is assigned a weight representing its importance to explain the phenotype.
The established link between kernels and pathways enables PIMKL to identify which molecular mechanisms are important for the prediction of the considered phenotype, as shown in Figure \ref{fig:pimkl_concept}. 

\subsection{Pathway Induction}
\label{ssec:pathway_induction}

PIMKL encodes information from the topology of each pathway's sub-network.
The approach of integrating pathway information into \textit{interaction-aware} kernel similarity functions is here termed \textit{pathway induction}.
Specifically, we design kernel functions by utilizing a positive semidefinite  (PSD) matrix that encodes the topological properties of a graph. Given any PSD matrix $M$, a valid kernel can be \textit{induced} through the following weighted inner product~\cite{M.Bishop2006}:
\begin{align*}
k(x,y) = x^{T} M y  \, .
\end{align*}
Hence, in order to have a \textit{pathway-induced} kernel, we only need to define $M$ such as it encodes the graph topological information of the pathway. 

In this work, an encoding based on the symmetric normalized Laplacian matrix is adopted.
Pathways are defined as weighted undirected graphs $\mathcal{P} = (V, E, W)$, with $N_{v}=|V|$ nodes, $N_{e}=|E|$ edges and a diagonal weight matrix $W \in \mathbb{R}^{N_{e} \times N_{e}}$, representing respectively the molecular entities, e.g. genes or proteins, their interactions and the weights associated with them.

For any pair of samples $x,y \in \mathbb{R}^{N_{v}}$, we define a \textit{pathway-induced} kernel using the  following similarity function:
\begin{align*}
\begin{cases}
&k_{\mathcal{L}}(x,y) =x^{T}\mathcal{L}y = x^{T}\mathcal{S}\mathcal{S}^{T}y = \Pi(x)^{T}\Pi(y)\\
&\mathcal{S} = D^{-\frac{1}{2}}SW^{\frac{1}{2}}
\end{cases}
\end{align*}
where $\mathcal{L} \in \mathbb{R}^{N_{v} \times N_{v}}$, $D \in \mathbb{R}^{N_{v} \times N_{v}}$ and $S \in \mathbb{R}^{N_{v} \times N_{e}}$ are respectively the normalized Laplacian, the diagonal degree matrix and an ordered incidence matrix for graph $\mathcal{P}$ associated with a pathway (see Supplementary \ref{sup:pathway_induction} for a detailed explanation about the formulation and the design of the kernel function).

\begin{wrapfigure}[32]{0}{0.4\textwidth}
\centering
\includegraphics[width=0.4\textwidth, clip]{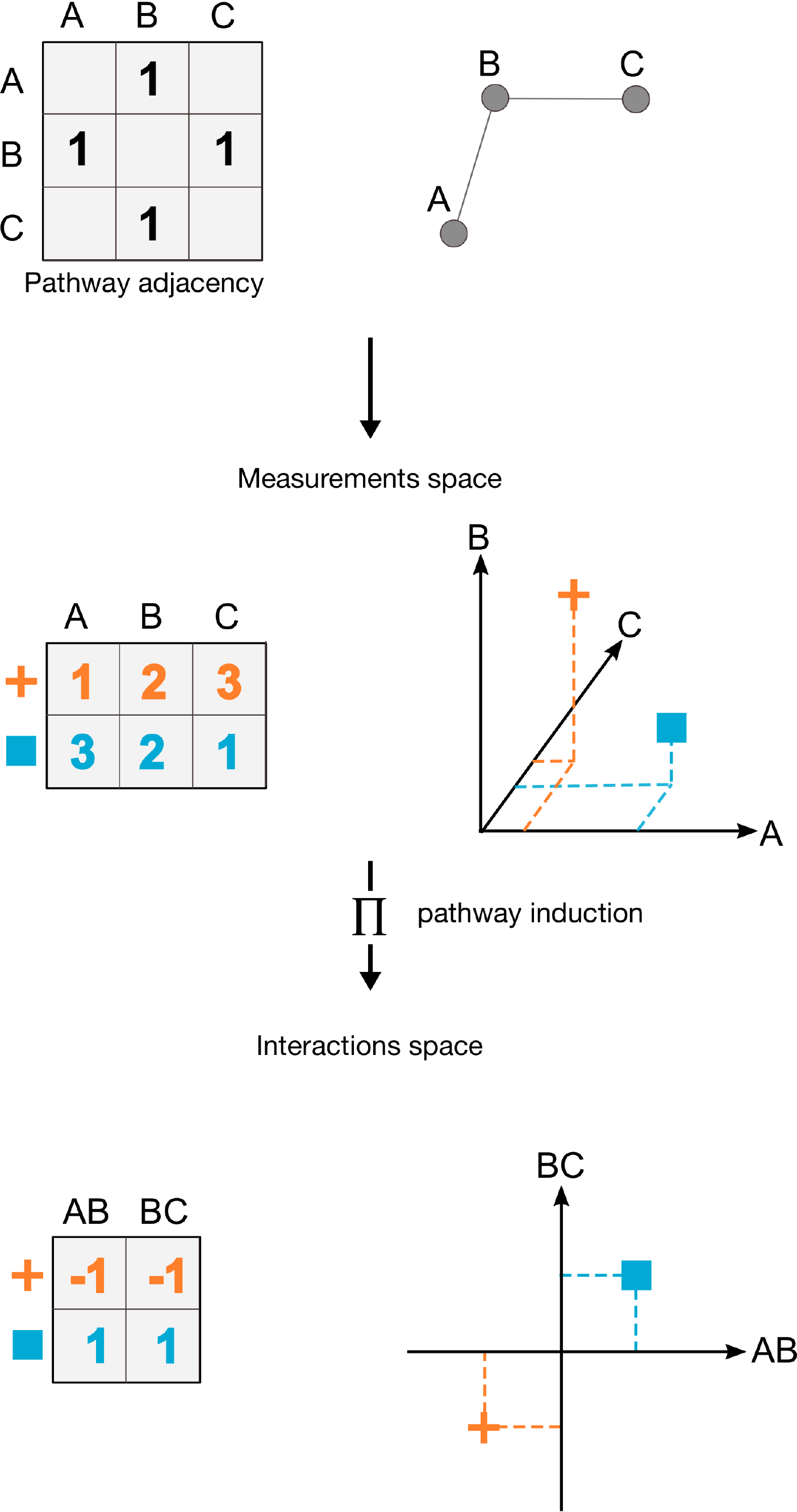}\\
\caption{
\textbf{Pathway induction.}
Given a pathway adjacency matrix, it is possible to map sample measurements from their original space, the space of the nodes, to the space of the interactions between the molecular entities. The example above shows how the mapping using \textit{pathway induction} transforms the considered samples.
}
\label{fig:pathway_induction}
\end{wrapfigure}

The normalized Laplacian can be interpreted as a discrete Laplace operator representing a diffusion process over graph nodes. A  \textit{pathway induction} process  based on it introduces a mapping $\Pi$ from the original space of the $N_{v}$ molecular measurements to an $N_{e}$-dimensional feature space, where each pathway interaction is a dimension, and the value along the edge is the discrete diffusion potential between the nodes that the edge connects.
A schematic illustration of the mapping introduced using \textit{pathway induction} can be seen in Figure \ref{fig:pathway_induction}.

\subsection{Pathway Induced Multiple Kernel Learning}
\label{ssec:pimkl}

PIMKL makes use of the concept of \textit{pathway induction}, defined in \ref{ssec:pathway_induction}, to implement a multiple kernel learning classification system.
Consider a network that recapitulates a comprehensive set of known molecular interactions represented by a graph $\mathcal{G}=(V, E, W)$ with $N_{v}=|V|$ nodes,  $N_{e}=|E|$ edges and a set of molecular measurements $X \in \mathbb{R}^{N \times N_{v}}$  with associated labels for a relevant phenotype $y$.

Given a selection of pathways $P$, e.g. gene sets from ontologies or inferred via community detection, it is possible to extract for each pathway $p \in P$, a corresponding sub-graph $\mathcal{P}^{p}=(V^{p}, E^{p}, W^{p}) \subset \mathcal{G}$ with  $N^p_{v}=|V^{p}|$ nodes, $N^{p}_{e}=|E^{p}|$ edges and a sub-selection of measurements corresponding to the genes contained in the pathway $X^{p} \in \mathbb{R}^{N \times N^{p}_{v}}$.

For every pathway, a Gram matrix $K^{p}$ can be used to represent the pathway-induced kernel, where $K^{p}$   is computed for each pair of samples $i$ and $j$ as follows:
\begin{align*}
K^{p}_{ij} =  k_{\mathcal{L}^{p}}(x_{i},x_{j})  \, .
\end{align*}
In the above equation,  $x_{i}, x_{j} \in \mathbb{R}^{N^{p}_{v}}$ and $\mathcal{L}^{p}$ is the normalized Laplacian for $\mathcal{P}^{p}\ \forall p \in P$.

For the problem of finding the optimal mixture of kernels over the different pathway-induced kernels, any supervised MKL algorithm can be used. In this work, a custom version of EasyMKL \cite{Aiolli2015} was implemented as it achieves high performance at a low  computational cost. EasyMKL is based on the Kernel method for the Optimization of the Margin Distribution (KOMD) \cite{Aiolli2008} and focuses on optimizing a linear combination of kernels:
\begin{align*}
K &=  \sum_{p=1}^{P} w_{p} K^{p}, \quad 
w_{p} \geq 0
\end{align*}
In PIMKL, the weights obtained are divided by their sum, as we are interested in evaluating the relative contribution of each kernel. This normalization does not affect the quality of the kernel mixture, which is invariant under positive scalar multiplication. In addition, to account for differences in sub-graph sizes we force the kernel matrices to have equal trace, ensuring comparable Gram matrices between different pathways.

It is important to note that PIMKL formulation enables a seamless integration of multi-omics data. Kernels from different data types can be easily generated and added to the mixture. The same applies to multi-modal data integration: kernels generated from other data modalities associated with a specific sample, e.g. histopathology images or clinical records, can be added to the mixture and weighted accordingly to their contribution in the classification problem.

\section{Results}
\label{sec:results}

In the following sections, we discuss the application of PIMKL to different breast cancer cohorts. First, in Section~\ref{ssec:microarray_cunetal}, PIMKL is compared to a previous study by Cun and Frölich~\cite{Cun2012} where different algorithms for phenotype prediction and gene selection using prior knowledge were compared. 
Later, in Section~\ref{ssec:metabric}, PIMKL is applied to gene expression and copy number data from the  METABRIC cohort~\cite{Curtis2012} with two purposes: first, we aim to test whether transfer learning between different studies is possible, and, second, we want to evaluate  PIMKL performance in the analysis of multi-omics analysis in the presence of noise or uninformative data.

\subsection{PIMKL on breast cancer microarray cohorts}
\label{ssec:microarray_cunetal}

PIMKL was tested on microarray gene expression data  from six breast cancer cohorts (see Supplementary Table \ref{tab:microarray_description} for details about the cohorts).
The classification task consisted in stratifying breast cancer samples according to occurrence of relapse within 5 years.
To ensure the fairest possible comparison, we used  the same interaction sources as in the study by Cun and Fröhlich, namely a merge between KEGG pathways and Pathway Commons. As access to the older release of KEGG is restricted, the most recent versions from both sources were used.
A collection of 50 \textit{hallmark} gene sets from the Molecular Signatures Database (MSigDB) version 5.2 \cite{Liberzon2015} was used to define the sub-graphs used for \textit{pathway induction}, generating $P=$ 50 kernels.
The classification performance was evaluated by means of the Area Under the receiver operating characterstic Curve (AUC). We closely followed the same data processing procedures and the cross-validation scheme as proposed in the original study  (for details, see Supplementary Algorithm \ref{sup:cunetal_cv}).

\begin{figure}[!b]
\centering
\begin{subfigure}[b]{0.45\textwidth}
\centering
\includegraphics[width=\textwidth]{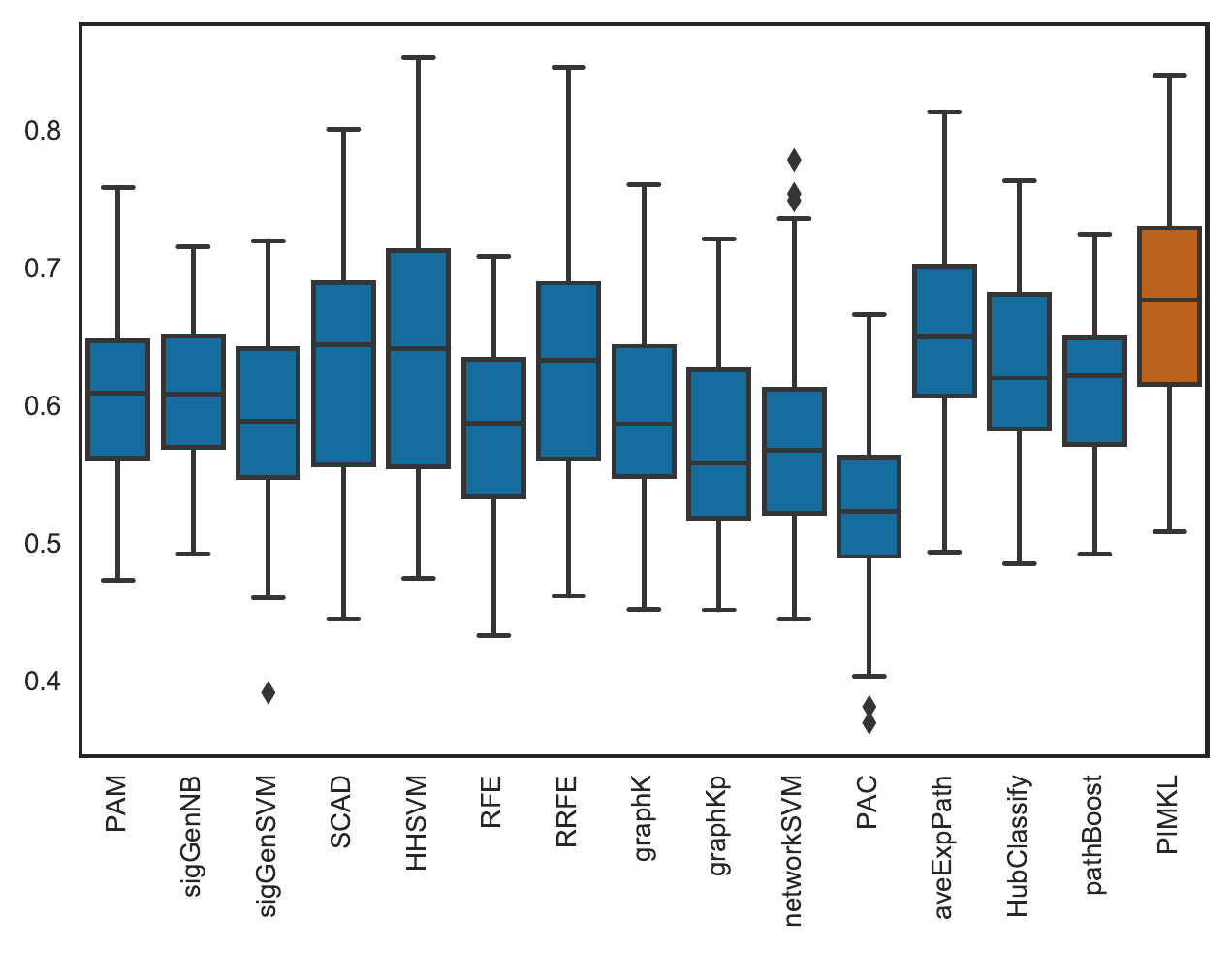}
\caption{\label{fig:pimkl_cunetal_aucs}}
\end{subfigure}
\begin{subfigure}[b]{0.54\textwidth}
\centering
\includegraphics[width=\textwidth]{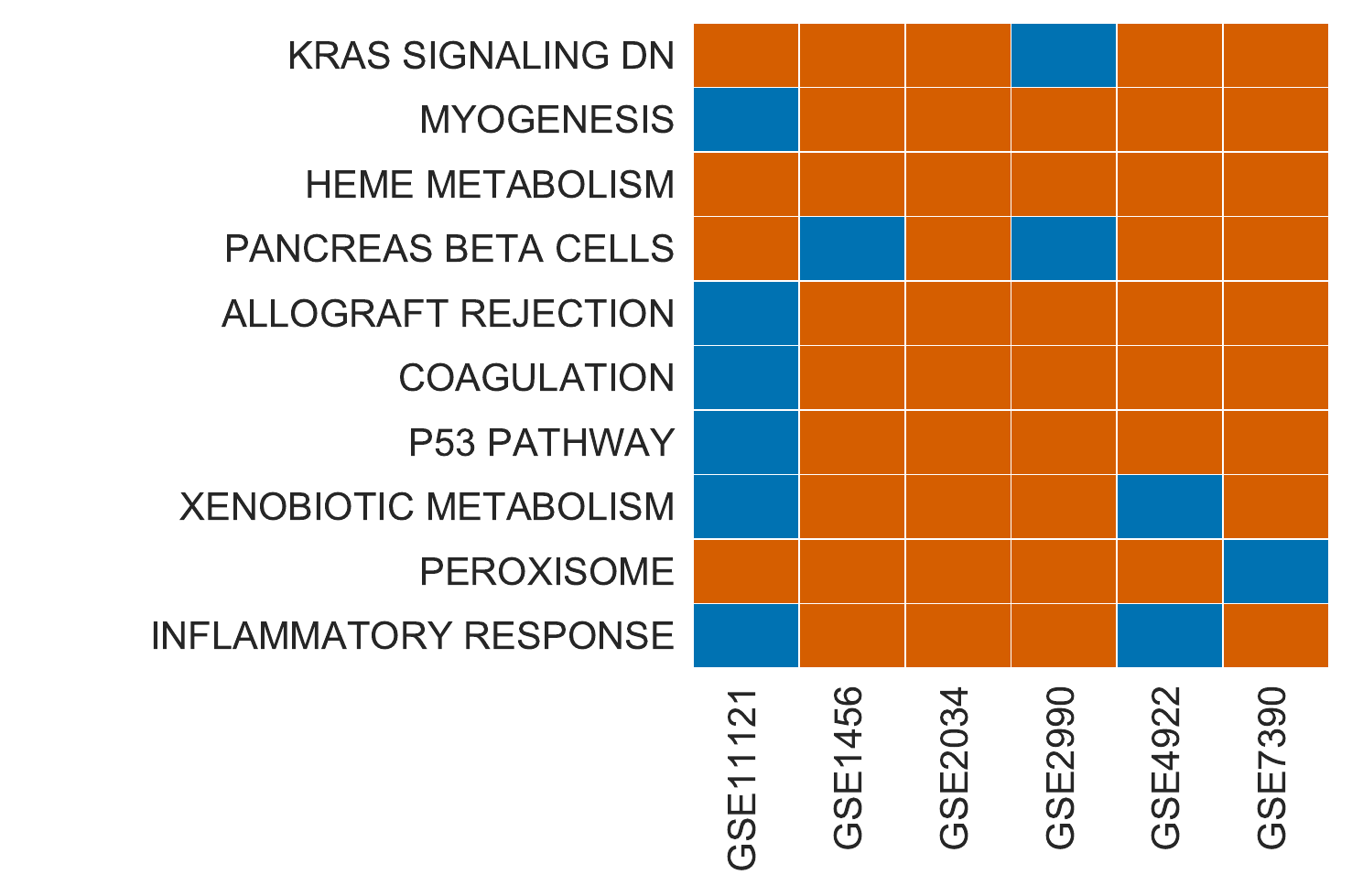}
\caption{\label{fig:pimkl_cunetal_weights}}
\end{subfigure}
\caption{
\label{fig:pimkl_cunetal}
\textbf{ PIMKL cross-validation results.}
(a) Box plots for AUC values over all cohorts for the methods considered. PIMKL results are reported in red, while other methods' results are colored in blue.
Box plots are obtained from ten (repeats of) mean AUC values over 10-fold cross-validation splits, see algorithm \ref{sup:cunetal_cv}.
(b) Heat map showing significant pathways selected by PIMKL across the different cohorts considered in the study. Significant pathways are highlighted in red, while non-significant are colored in blue.
} 
\end{figure}

The results of PIMKL compared to the 14 algorithms considered by Cun and Fröhlich are reported in Figure \ref{fig:pimkl_cunetal}.
Overall AUC values for the 6 cohorts over the cross-validation rounds for all considered methods are shown in Figure \ref{fig:pimkl_cunetal_aucs}. 
AUC values for the single cohorts can be found in Supplementary Figure \ref{sup:pimkl_cunetal_cv_cohorts}, where PIMKL exhibits the highest median value and  consistently outperforms the other methods or is in the top performers group on single cohorts.

Results are consistent when other gene sets are used, even when we use randomized versions of  functionally related gene sets  (see Supplementary Figure \ref{sup:pimkl_cunetal_cv_genesets}). These results prove that PIMKL performance does not depend on the specific selection of pathways, and that through the MKL optimization we can identify the informative gene sets in disparate collections of genes. Notice, however, that while choosing random gene sets does not worsen PIMKL performance,  interpretation of the molecular signatures, as we will discuss next, is only possible when the sets have a well defined biological function. 

As discussed in Section \ref{sec:methods}, PIMKL generates a molecular signature  given by the weighted contributions of each kernels. Each weight represents the relative importance of  each hallmark pathway used for \textit{pathway induction} to explain the phenotype. To evaluate the stability of the signature, the pathway weight distribution over cross-validation rounds was analyzed. Our baseline is the case where all kernels have the same weight: $w_b = \frac{1}{P}$, representing a situation where no pathway contributes more than the others to the phenotype prediction. 
To find whether a pathway is  significant for the phenotype, 
the distribution of the kernel weights with median above $w_b$ was tested against the baseline using a one-sample Wilcoxon signed-rank test.
$p$-values at significance level 0.001 were corrected for multiple testing using Benjamini-Hochberg (for details see Supplementary Figure \ref{sup:pimkl_cunetal_cv_cohorts_weights}).
Pathways where the significance was achieved in at least four of six cohorts are reported in Figure \ref{fig:pimkl_cunetal_weights}.

Interestingly, heme metabolism pathway is significant in all cohorts. This pathway is involved in the metabolism of heme and erythroblast differentiation. A possible explanation is that heme metabolism might reflect an active vascularization of the samples, a phenomenon widely observed in cancer progression \cite{pmid17717633}.
A more intriguing hypothesis is a possible association between elevated heme metabolism and cancer progression, as has been reported in non-small-cell lung cancer cells and xenograft tumors \cite{hooda2015evaluating}.
It is also interesting to look at the pathways that are significant in at least five cohorts: KRAS signaling, myogenesis, allograft rejection, coagulation, P53 pathway and peroxisome. 
All of these pathways are associated with breast cancer. For instance, activation of KRAS signaling has been reported to promote the mesenchymal features of basal-type breast cancer \cite{pmid25633745, Najumudeen2016}. Myogenesis, or the process of formation of muscular tissue, is commonly disrupted in cancer \cite{pmid29153940}. Allograft rejection might reflect an immune-mediated tumour rejection signature following administration of immunotherapeutic agents \cite{Bedognetti2015}. Several studies have suggested a role for blood coagulation proteins in tumour progression \cite{Lima2013, pmid15905465, FALANGA2013}. P53 is the most commonly mutated protein in cancer \cite{Vazquez2008,Mandinova2011}. Finally, peroxisomes are small, membrane-enclosed organelles that contain enzymes involved in a variety of metabolic reactions, including several aspects of energy metabolism. Altered peroxisome metabolism has been linked to various diseases, including cancer \cite{pmid23674998, Fransen2012}.

\begin{figure}[!htb]
\centering
\includegraphics[width=0.7\textwidth, clip]{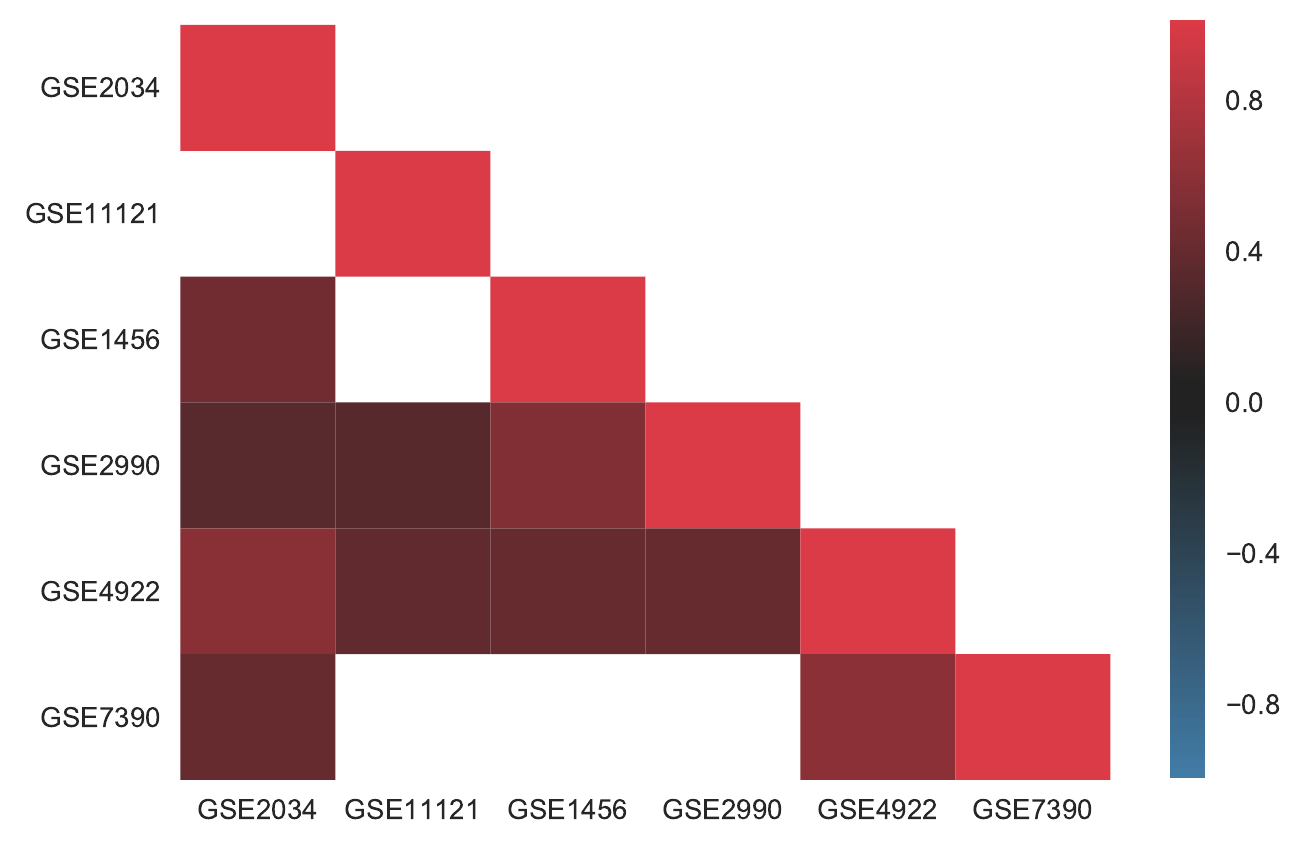}\\
\caption{
\label{fig:pimkl_cunetal_cv_cohorts_weights_correlation}
\textbf{ Correlation in molecular signatures.}
Heat map reporting the correlation of the molecular signature estimated across multiple cohorts.
Correlation values are reported in the lower triangular part of the heat map (since it is symmetric) on blue to red scale, white squares indicate non significant correlations.
All cohorts exhibit a positive correlation, significant in most cases, proving the stability of the molecular signature obtained with PIMKL.
} 
\end{figure}

Finally, Figure \ref{fig:pimkl_cunetal_cv_cohorts_weights_correlation} reports the correlation of the PIMKL molecular
signatures estimated across multiple cohorts and highlights their stability across different studies,  suggesting that a cohort-independent  disease free survival  signature for breast cancer has been learned.

\subsection{PIMKL on METABRIC cohort}
\label{ssec:metabric}

To test PIMKL  applicability to multi-modal datasets,  we used our methodology to 
predict disease free survival in
the  METABRIC  breast cancer cohort, consisting of 1890 samples profiled with Illumina Human v3 microarray data (mRNA) and Affymetrix SNP 6.0 copy number data (CNA), see Supplementary Table \ref{tab:metabric_description} for details.

\begin{figure}[!htb]
\centering
\includegraphics[width=0.7\textwidth, clip]{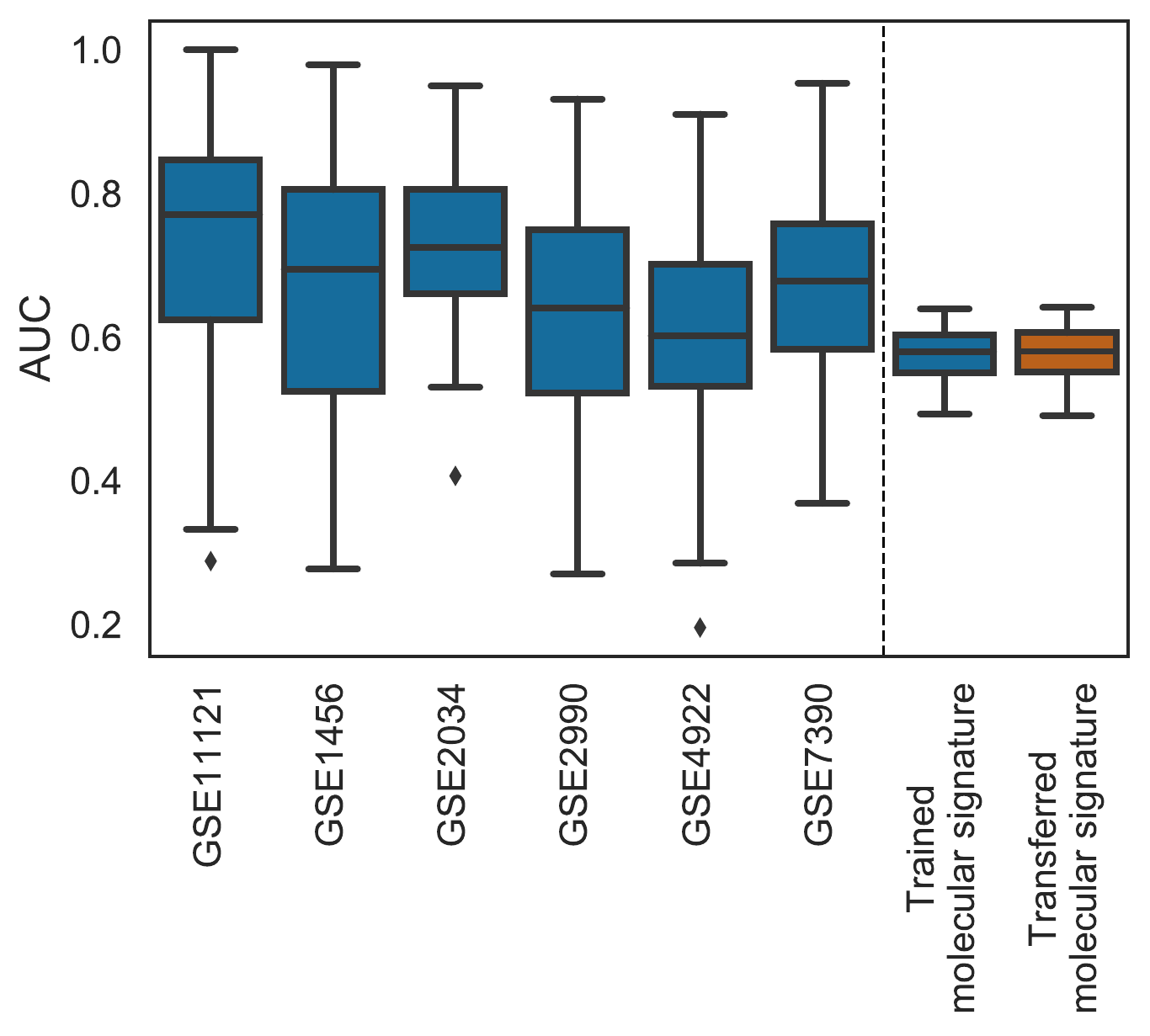}\\
\caption{
\label{fig:pimkl_metabric_transfer_learning}
\textbf{ PIMKL performance on METABRIC.}
Box plots of the performance of PIMKL over the six cohorts used to benchmark the method (left of the dashed vertical line) and its application on METABRIC for disease free survival prediction (right of the dashed vertical line). Optimized weights at training by EasyMKL (blue); provided weights from taking the pathway-wise median weights of the six signatures obtained during benchmarking (red).
} 
\end{figure}

\begin{figure}[!htb]
\centering
\includegraphics[width=0.7\linewidth, clip]{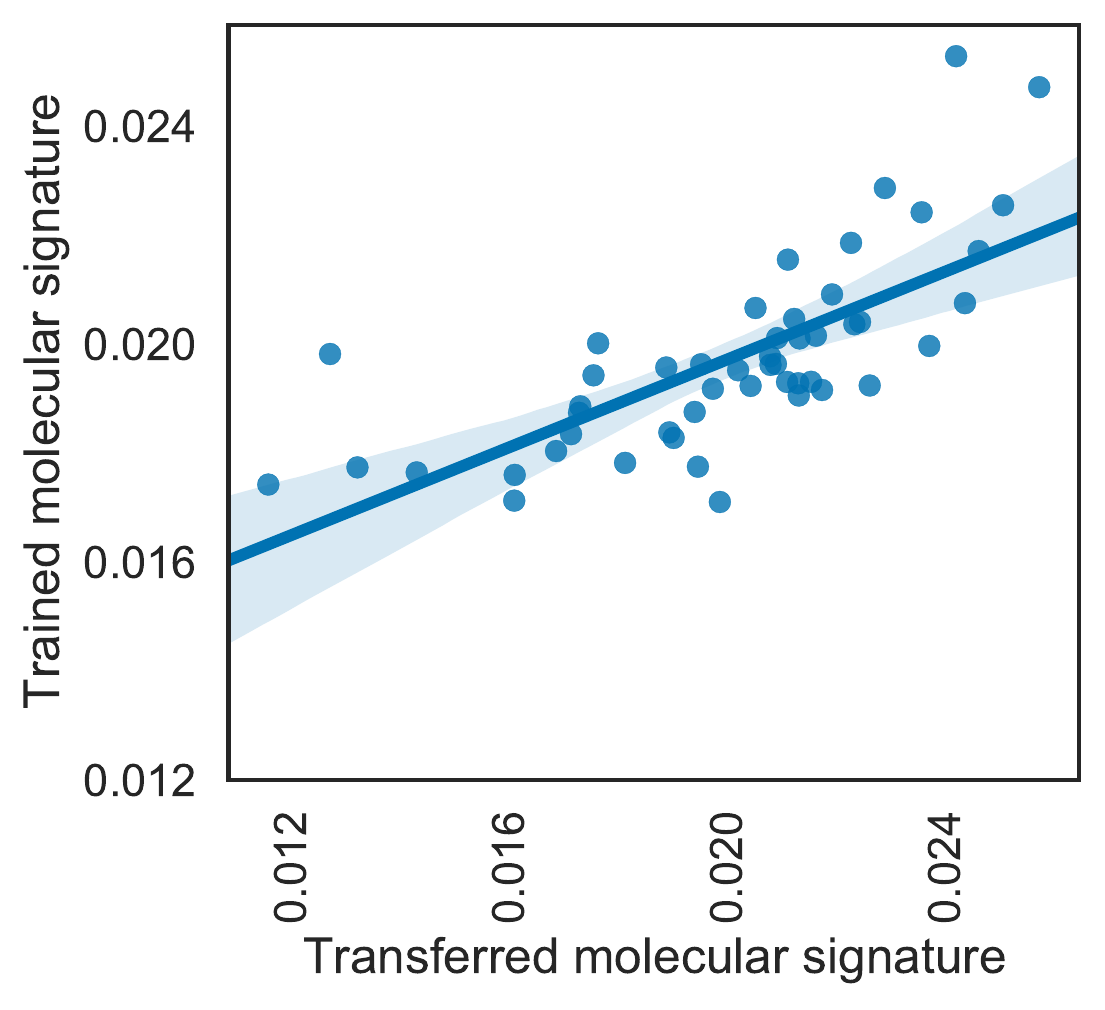}\\
\caption{
\label{fig:pimkl_transfer_weights_regression}
\textbf{ Regression between trained and transferred signature.}
Regression of the pathway weights of the signature obtained from directly training on METABRIC (median over 100 cross-validation folds) against the transferred signature obtained from training on six independent cohorts  (each median over 100 cross-validation folds) indicating high correlation of the two signatures.
} 
\end{figure}

In order to validate the generalization power of PIMKL-generated molecular signatures, we first focused on the analysis of METABRIC microarray data. Our hypothesis here is that the underlying  molecular mechanisms associated with  disease free survival are the same in different cohorts and, as such, knowledge learned in one cohort can be transferred to another one. 
After computing the \textit{pathway-induced} kernels with the same procedure adopted in Section~\ref{ssec:microarray_cunetal}, a set of pathway weights was defined using the median of the weights obtained in the six previously analyzed cohorts.
Figure  \ref{fig:pimkl_metabric_transfer_learning}  shows the results obtained by training a KOMD classifier using the weights transferred from the six independent cohorts and by learning METABRIC-specific pathway weights  (for details see Supplementary Algorithm \ref{sup:metabric_cv}). It is evident that both molecular signatures perform  very similarly. Indeed, the two signatures are highly correlated (Pearson correlation $\rho=0.72$, $p$-value $=3.34 \cdot 10^{-9}$, Figure \ref{fig:pimkl_transfer_weights_regression}). It is important to notice that the variance of the prediction results is also consistently reduced, probably due to the newer microarray technology used by the METABRIC study.

To  test PIMKL capability to integrate multi-omics data, both the mRNA and CNA data from the METABRIC cohort were jointly utilized in the same predictive task. A set of additional kernels were generated using the copy number data and then used in two ways: First, the CNA kernels were independently optimized with PIMKL, and second, a mixture of CNA and mRNA kernels were jointly optimized.

\begin{figure}[!htb]
\centering
\includegraphics[width=0.7\textwidth]{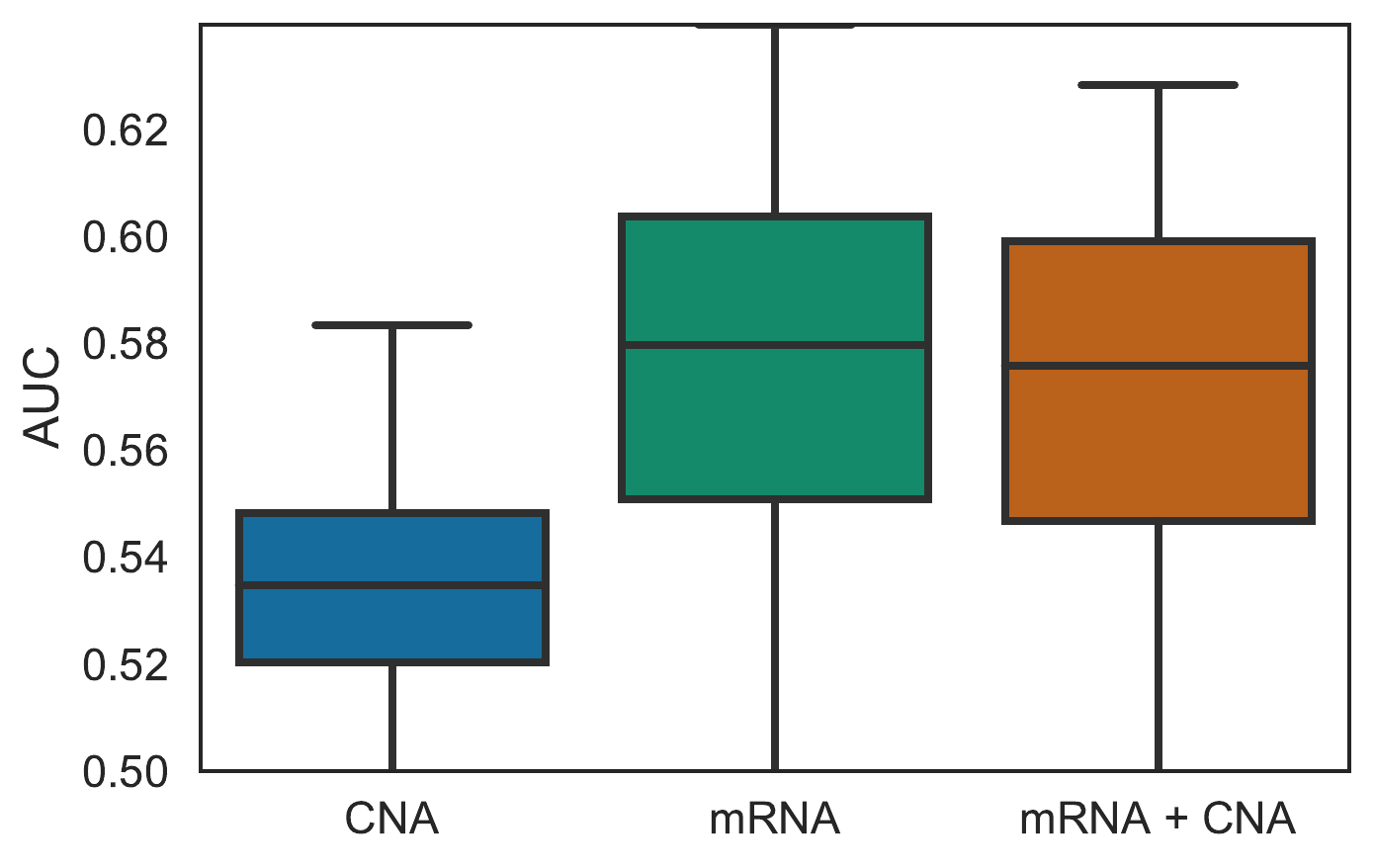}\\
\caption{
\label{fig:pimkl_metabric_multiomics}
\textbf{ PIMKL performance on METABRIC multi-omics.}
Box plots for AUC values obtained applying PIMKL on different data types and their integration. CNA only results are reported in blue, mRNA ones in green and their integration in red.
}
\end{figure}

From Figure \ref{fig:pimkl_metabric_multiomics}, it is evident that the   CNA data are not as predictive as mRNA  regarding disease free survival. However,  it is interesting to notice that PIMKL is able to discard noisy kernels -- associated with CNA data -- to achieve  similar levels of performance  when using the more informative mRNA data and when using a mixture of CNA and mRNA data. This suggests that the application of the proposed algorithm is feasible even  when no prior knowledge about the information content of each single omic type is available.

\section{Discussion}

We have presented here PIMKL (Pathway Induced Multiple Kernel Learning), a novel, effective and interpretable machine learning methodology for phenotype prediction using multi-modal molecular data.
PIMKL is based on a multiple kernel learning (MKL) framework, a kernel-based method that has demonstrated excellent capabilities to integrate multi-omics  datasets~\cite{Costello2014}.
In addition, PIMKL also exploits prior knowledge in the form of molecular interaction networks and sets of annotated pathways with known biological functions to build a mixture of \textit{pathway-induced} kernels.
The main novelty introduced in this work is the definition of multiple interaction-aware kernel functions, which enables us to encode information about the molecular prior knowledge related to a phenotype, and facilitates the interpretation of the results in terms of known biological functions and/or  specific molecular interactions. We achieve this by using such kernels to map samples into the space of network edges, i.e. molecular interactions, recovering a direct biological interpretation.
The kernel weights are later optimized  to classify a  phenotype or a clinical variable of interest.

In this work, PIMKL was extensively tested in the context of  predicting disease-free survival from breast cancer samples. We have demonstrated that the resulting weighted combination of kernels can be interpreted as a phenotypic molecular signature and provides  insights into the underlying molecular mechanisms. 
As a benchmark, a well-studied set of cohorts previously analyzed using a range of stratification methods has been adopted~\cite{Cun2012}. The quality and the stability of the obtained signatures has been thoroughly investigated, and we have shown  that PIMKL outperforms other methods and finds stable molecular signatures across different breast cancer cohorts.
We also investigated the generalization power of the found signatures by testing them  on unseen mRNA breast cancer data from the METABRIC cohort and the associated  disease-free survival data. The obtained results  have confirmed that the algorithm can be used to effectively gain insights into disease progression and that this knowledge can be transferred to other cohorts without loss of performance. 

Furthermore, PIMKL can be seamlessly applied to integrate data from different omic layers. Its intrinsic capability to discard noisy molecular features has been demonstrated by applying it on METABRIC, where it was possible to integrate multiple types of data with varying predictive power. Even when non-informative data was mixed with informative data, PIMKL was able to discard uninformative kernels and achieve similar levels of performance.
Evidently, PIMKL is not restricted to breast cancer, the specific omic data types or the sources of prior information used in this work. Its application is open to other disease types using any available combination of data together with any suitable prior network and sets of genes. 

Besides being capable of using different types of prior knowledge, the proposed approach is also highly flexible with regard to the number and nature of the selected kernels.
Indeed, PIMKL was developed by making use of an efficient implementation of EasyMKL~\cite{Aiolli2015}, an extremely scalable MKL algorithm with constant memory complexity in the number of kernels. 
This efficiency can potentially allow the user to define smaller pathways, leading to a more fine-grained characterization and understanding of the molecular mechanisms involved in disease progression with limited performance drawbacks.

Finally, possible extensions of PIMKL, such as optimizing the kernel mixture using semi-supervised or unsupervised multiple kernel learning methodologies \cite{Mariette2017}, may help  discovery  phenotype-independent pathway signatures and will be explored in the future.
To summarize, PIMKL provides a flexible and scalable method to translate prior knowledge and molecular data into actionable insights in a clinical setting. 

\section*{Declarations}

\subsection*{Acknowledgments}
\begin{figure}[!htb]
\centering
\includegraphics[width=0.7\textwidth, clip]{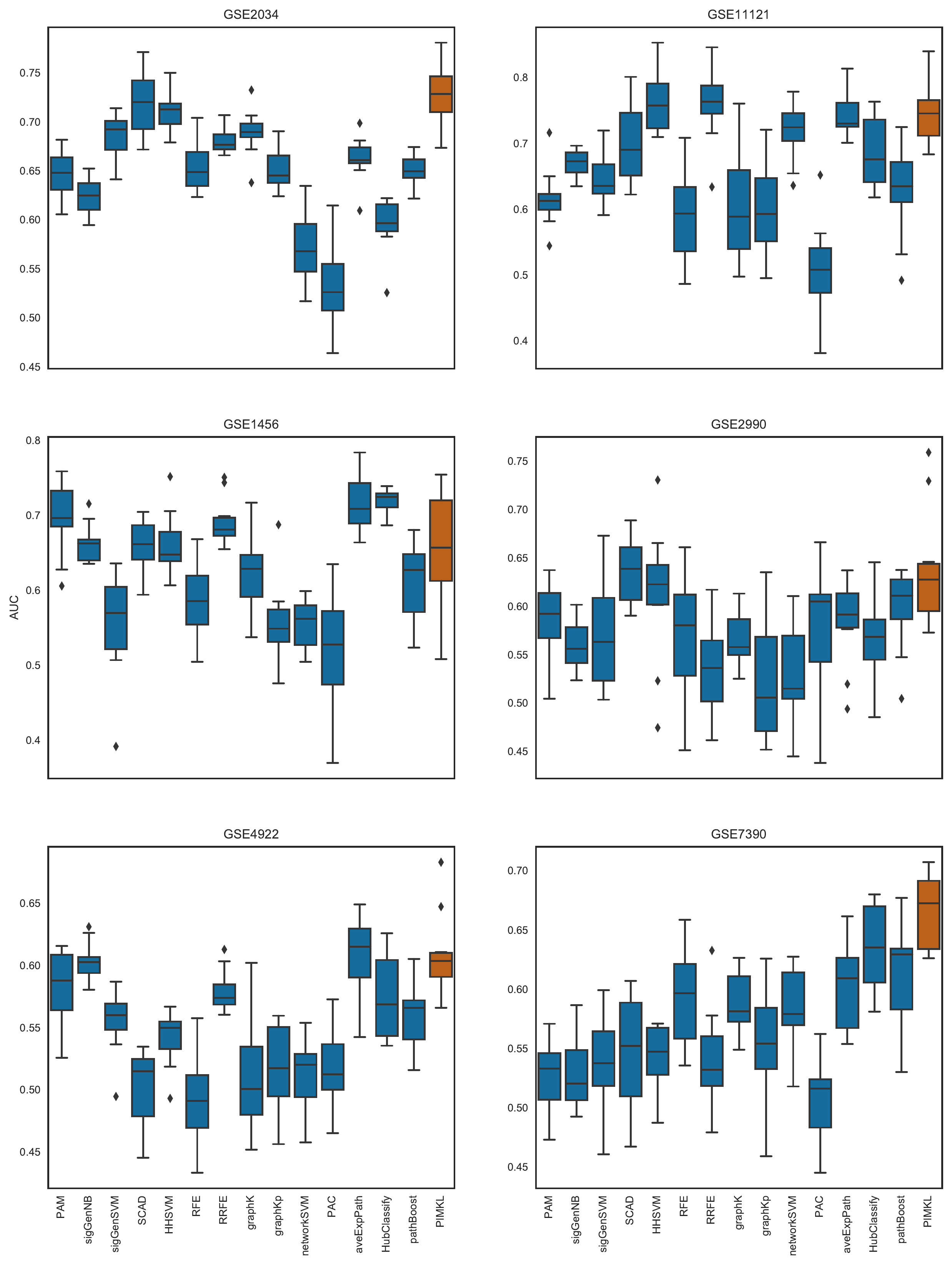}\\
\caption{
\label{sup:pimkl_cunetal_cv_cohorts}
\textbf{ PIMKL cross-validation AUC.}
Box plots of the AUC values for the methods analyzed in \cite{Cun2012} (blue) and PIMKL (red). PIMKL clearly outperforms other methods in four out of six data sets. For GSE1456 is performing close to other methods average while for GSE11121 is in the top group.
Results are presented as in \cite{Cun2012}, where each box is drawn from ten (repeats of) mean AUC values over 10-fold cross-validation splits, see algorithm \ref{sup:cunetal_cv}.
} 
\end{figure}

The project leading to this application has received funding from the European Union’s Horizon 2020 research and innovation program under grant agreement No 668858.
We thank Yupeng Cun for providing results \cite{Cun2012} for recreation of Figures \ref{sup:pimkl_cunetal_cv_cohorts} and \ref{fig:pimkl_cunetal_aucs}.

\subsection*{Authors contributions}

M.M., J.C., R.M. and M.R.M. conceived the study and analyses. M.M., J.C. and R.M. implemented PIMKL and performed data analysis. M.R.M. provided  biological analysis and interpretation.
M.M., J.C., R.M. and M.R.M. wrote the manuscript with input from all authors. 

\subsection*{Competing financial interests} 

The authors declare no competing financial interest.

\subsection*{Availability of data and materials}

Processed data and materials used to produce the results presented in this work can be downloaded from the following link \url{https://ibm.box.com/s/ac2ilhyn7xjj27r0xiwtom4crccuobst}. PIMKL as a service is available on IBM Cloud at the following link \url{https://sysbio.uk-south.containers.mybluemix.net/pimkl/}. A set of anonymous credentials can be created for reviewers.

\bibliographystyle{plainnat}
\setlength{\bibsep}{0pt plus 0.3ex}

\begin{footnotesize}
\bibliography{main}
\end{footnotesize}
\newpage

\beginsupplement

\begin{Large}
\noindent \textbf{PIMKL: Pathway Induced Multiple Kernel Learning}
\end{Large}\\
\begin{footnotesize}
\noindent Matteo Manica\textsuperscript{1,2,*}, Joris Cadow \textsuperscript{1,2,*}, Roland Mathis \textsuperscript{1,*},  María Rodríguez Martínez\textsuperscript{1, $\dagger$}\\
\noindent \textit{ \{tte, dow, lth, mrm\}@zurich.ibm.com}\\
\textsuperscript{1} IBM Research Zürich\\
\textsuperscript{2}	ETH - Zürich\\ 
\textsuperscript{*} Shared first authorship \\
\textsuperscript{ $\dagger$} Corresponding author
\end{footnotesize}

\section*{Supplementary information}

\subsection{Pathway induction}
\label{sup:pathway_induction}

Similarity functions can be designed by making use of a PSD matrix to \textit{induce} a weighted inner product:
\begin{align*}
k(x,y) = x^{T} M y\quad \forall x,y \in \mathbb{R}^{N}
\end{align*}
represents a valid kernel if matrix $M \in \mathbb{R}^{N\times N}$ is PSD~\cite{M.Bishop2006}, indeed this ensures the existence of a matrix $U$:
\begin{align*}
M &= U^{T} U\\
\phi(x) &= Ux
\end{align*}
where $\phi$ is a mapping describing a transformation in the feature space.

By making use of a PSD matrix encoding the topological properties of a graph representing a pathway, it's possible to design \textit{interaction-aware} kernels.

Let's consider an undirected graph representing a pathway:
\begin{align*}
\mathcal{P} = (V, E)
\end{align*}
with $N_{v}=|V|$ nodes and  $N_{e}=|E|$ edges representing the genes/proteins and their interactions respectively.
Such a graph is defined by a symmetric adjacency matrix $A \in \{0,1\}^{N_{v} \times N_{v}}$:
\begin{align*}
A_{ij} = 1\quad \forall (i,j) \in E
\end{align*}
and its diagonal degree matrix $D \in \mathbb{R}^{N_{v} \times N_{v}}$:
\begin{align*}
D_{jj} = \sum_{i}A_{ij}
\end{align*}
For such a graph, the Laplacian matrix $L \in \mathbb{R}^{N_{v} \times N_{v}}$ is computed using the following:
\begin{align*}
L = D - A
\end{align*}
The Laplacian is a PSD matrix and therefore represents a suitable candidate for \textit{induction} of a weighted inner product based on a pathway topology. This can be shown by defining an ordered incidence matrix $S \in \mathbb{R}^{N_{v} \times N_{e}}$ for $\mathcal{P}$ that by construction satisfies the relation $L=SS^{T}$.
As in \cite{Anderson1985}, after introducing an index set $\mathcal{E}$ for the edges $E$, $S$ can be defined as:
\begin{align*}
S_{ne} &= 
\begin{cases}
1\quad &\text{if}\ n=i \wedge i \leq j\\
-1\quad &\text{if}\ n=j\\
0\quad &\text{otherwise}
\end{cases}
\\
\text{where}\ e \in \mathcal{E}\ &\text{corresponds to edge}\ (i,j) \in E\ \text{and}\ n \in V
\end{align*}


Moreover, the Laplacian can be interpreted as a discrete Laplace operator. Indicating with $X \in \mathbb{R}^{N \times N_{v}}$ a set of $N$ samples, a discrete diffusion process over graph nodes is described as:
\begin{align}
\label{al:laplace_diffusion}
LX^{T}=SS^{T}X^{T}
\end{align}
where the term $S^{T}X^{T}$ computes the discrete diffusion potential (a difference) along the edges and Equation \ref{al:laplace_diffusion} describes how the flow of this potential is effected, aggregating incoming and outgoing flows on the nodes.

Decomposing the Laplacian using an ordered incidence matrix shows how samples $X$ are mapped from the original space with measurements for $N_{v}$ molecular entities into an $N_{e}$-dimensional feature space, where each interaction from the pathway is a dimension and the value along the edge is the discrete diffusion potential between respective node's measurements.

The inner product in this space is the resulting similarity function defined as: 
\begin{align*}
k_{L}(x,y) = x^{T}Ly = x^{T}SS^{T}y\quad \forall x,y \in \mathbb{R}^{N_{v}}
\end{align*}

Similar considerations can be applied to weighted graphs with non-negative weights. Given a weighted undirected graph $\mathcal{P} = (V, E, W)$, indicating by $W \in \mathbb{R}^{N_{e} \times N_{e}}$ its diagonal weights matrix, the Laplacian $L$ is defined as:
\begin{align*}
L &= SWS^{T}\\
L_{ij} &=
\begin{cases}
d_{i} - W_{e} & \text{if}\ i=j\\
- W_{e} & \text{otherwise}
\end{cases}
\\
\text{where}\ e \in \mathcal{E}\ &\text{corresponds to edge}\ (i,j) \in E
\end{align*}

To ensure an equal contribution from all the nodes in the considered pathway, the degree-normalized version of the Laplacian $\mathcal{L}$ can  be adopted:
\begin{align*}
\mathcal{L} &= D^{-\frac{1}{2}}SWS^{T}D^{-\frac{1}{2}} \\
\mathcal{L}_{ij} &=
\begin{cases}
1 - \dfrac{W_{e}}{d_{i}} & \text{if}\ i=j\ \text{and}\ d_{i}\neq0 \\
- \dfrac{W_{e}}{\sqrt{d_{i}d_{j}}} & \text{if}\ i\ \text{and}\ j\ \text{are adjacent}\\
0 & \text{otherwise}
\end{cases}
\\
\text{where}\ e \in \mathcal{E}\ &\text{corresponds to edge}\ (i,j) \in E
\end{align*}

This pathway encoding directly leads to the definition of \textit{pathway induction} used in this work. Given any two samples measurement $x,y \in \mathbb{R}^{N_{v}}$:
\begin{align*}
k_{\mathcal{L}}(x,y) = &x^{T}\mathcal{L}y =\\ =&
x^{T}D^{-\frac{1}{2}}SWS^{T}D^{-\frac{1}{2}}y =
x^{T}(D^{-\frac{1}{2}}SW^{\frac{1}{2}})(W^{\frac{1}{2}}S^{T}D^{-\frac{1}{2}})y =\\
=&x^{T}\mathcal{S}\mathcal{S}^{T}y = \Pi(x)^{T}\Pi(y)
\end{align*}
with:
\begin{align*}
\Pi(x) &=
\begin{cases}
  \sqrt{W_{e}} \dfrac{x_{i}}{\sqrt{d_{i}}} & \text{if}\ i=j\ \text{and}\ d_{i}\neq0 \\
  \sqrt{W_{e}}( \dfrac{x_{i}}{\sqrt{d_{i}}} - \dfrac{x_{j}}{\sqrt{d_{j}}} ) & \text{if}\ i\ \text{and}\ j\ \text{are adjacent}\\
  0 & \text{otherwise}
\end{cases}
\\
\text{where}\ e \in \mathcal{E}\ &\text{corresponds to edge}\ (i,j) \in E
\end{align*}

A similar concept was proposed \cite{Chen2011} at the complete network level.
The normalized Laplacian was used as a regularizer to constrain the optimization problem when training an SVM.
In PIMKL, we arrive at a similar formulation of the problem by introducing a feature mapping instead of using the Laplacian as a regularizer. We define a kernel function which allows easy application to any kernelized method and any further kernel transformation (e.g: polynomial, Gaussian, etc.). The decomposition of $\mathcal{L}$ can be derived from the graph but is implicit, and can be easily extended to the multiple kernel learning case, allowing us to work at pathway/sub-network level.

It should be noted that in PIMKL the individual pathway-induced kernels are set to equal trace (equal average self similarity of the samples) to learn fair relative weights independent of the sub-network size.

\subsection{Breast cancer microarray cohorts}
\label{sup:cunetal_microarray}

\begin{table}[!htb]
\caption{
\textbf{Breast cancer benchmark cohorts.}
Brief description of sample counts in the different classes for the cohorts considered in \cite{Cun2012} (all Affymetrix Human Genome U133A Array). In GSE4922 and GSE11121 metastasis free survival (dmfs) is considered, in other cohorts relapse free survival (rfs).
}
\small
\begin{tabular}{llll}
\addlinespace
\toprule
\textbf{GEOid} \cite{Barrett2013}& \textbf{Patients} & \textbf{dmfs/rfs $\leq$ 5 years} & \textbf{dmfs/rfs $>$ 5 years} \\
\midrule
GSE2034 \cite{Wang2005}& 286 & 93 & 183  \\
GSE1456 \cite{Pawitan2005}& 159 & 34 & 119  \\
GSE2990 \cite{Sotiriou2006}& 187 & 42 & 116  \\
GSE4922 \cite{Ivshina2006}& 249 & 69 & 159  \\
GSE7390 \cite{Desmedt2007}& 198 & 56 & 135  \\
GSE11121 \cite{Schmidt2008}& 200 & 28 & 154  \\
\bottomrule
\end{tabular}
\label{tab:microarray_description}
\end{table}

\begin{table}[!htb]
\caption{
\textbf{Breast cancer METABRIC cohort.} Brief description of sample counts in the different classes for the considered data types in the METABRIC (Molecular Taxonomy of Breast Cancer International Consortium) cohort \cite{Curtis2012}.
}
\small
\begin{tabular}{llll}
\addlinespace
\toprule
\textbf{Data types} & \textbf{Patients} & \textbf{Recurred/Progressed} & \textbf{DiseaseFree} \\
\midrule
Illumina Human v3 microarray (mRNA) & 1890 & 647 & 1333  \\
Affymetrix SNP 6.0 copy number (CNA) &&&\\
\bottomrule
\end{tabular}
\label{tab:metabric_description}
\end{table}

\begin{algorithm}[!htb]
\caption{
\textbf{PIMKL Cross-validation on \cite{Cun2012}.}
Cross-validation analysis of PIMKL for each of the breast cancer cohorts as suggested in \cite{Cun2012} (with internal optimization of parameters). Given as input: $X$ gene expression measurements with related clinical labels $y$, a set of $P$ pathways and a set of hyper-parameters $\Lambda$ = \{0, 0.1, 0.3, 0.5, 0.7, 0.9, 1.0\} for EasyMKL.
}
\label{sup:cunetal_cv}
\begin{algorithmic}[1]
\For{$i\gets 1, 10$}
    \For{$(X_{train}, y_{train}), (X_{validation}, y_{validation})\gets$ stratified 10-fold split $X,y$}
        \State learn feature-wise normalization on $X_{train}$ and apply to $X_{validation}$ 
        \For{($X_{train}^{\lambda},y_{train}^{\lambda}), (X_{test}^{\lambda}, y_{test}^{\lambda})\gets$ stratified 3-fold split of $X_{train}, y_{train}$}
        \Comment parameter grid search with 3-fold cross-validation
            \For{$\lambda \in \Lambda$}
                \State train PIMKL($\lambda$) on $\{ k_{\mathcal{L}^{p}}(X_{train}^{\lambda},X_{train}^{\lambda}) : p \in P\}$ and $y_{train}^{\lambda}$
                \State record prediction accuracy on $\sum_{p=1}^{P} k_{\mathcal{L}^{p}}(X_{train}^{\lambda},X_{test}^{\lambda})$
            \EndFor
        \EndFor
        \State $\lambda^{*} \gets$ argmax(mean prediction accuracy over cross-validation)
        \State PIMKL($\lambda^{*}$) on $\{ k_{\mathcal{L}^{p}}(X_{train},X_{train}) : p \in P\}$ and $y_{train}$
        \State report kernel weights $\boldsymbol{w}$
        \State report area under the curve for prediction on $\sum_{p=1}^{P} w_{p}k_{\mathcal{L}^{p}}(X_{train},X_{validation})$
    \EndFor
    \State report mean area under the curve over 10-fold splits
    \Comment for figure \ref{sup:pimkl_cunetal_cv_cohorts} and \ref{fig:pimkl_cunetal_aucs}
\EndFor
\end{algorithmic}
\end{algorithm}

\begin{algorithm}[H]
\caption{
\textbf{PIMKL Cross-validation on METABRIC.}
Cross-Validation on METABRIC single omics or multi-omics. Given as input: $X$ molecular measurements comprised of a selection of data types $T$ (CNA, mRNA or both) with related clinical labels $y$, a set of $P$ pathways with a respective pathway for each data type in $T$ and $\lambda$ = 0.2 for EasyMKL.
}\label{sup:metabric_cv}
\begin{algorithmic}[1]
\For{100 folds with 20 samples per class in $(X_{train}, y_{train})$}
    \For{$type$ in $T$ } 
        \State learn feature-wise normalization on $X_{type, train}$ and apply to $X_{type, validation}$
        \EndFor
    \State train PIMKL($\lambda$) on $\{ k_{\mathcal{L}^{p}}(X_{train},X_{train}) : p \in P\}$ and $y_{train}$
    \State report kernel weights $\boldsymbol{w}$
    \State report area under the curve for prediction on $\sum_{p=1}^{P} w_{p}k_{\mathcal{L}^{p}}(X_{train},X_{validation})$
\EndFor
\end{algorithmic}
\end{algorithm}

\begin{figure}[!htb]
\centering
\includegraphics[width=0.7\textwidth, trim={0 0.7cm 0 0}, clip]{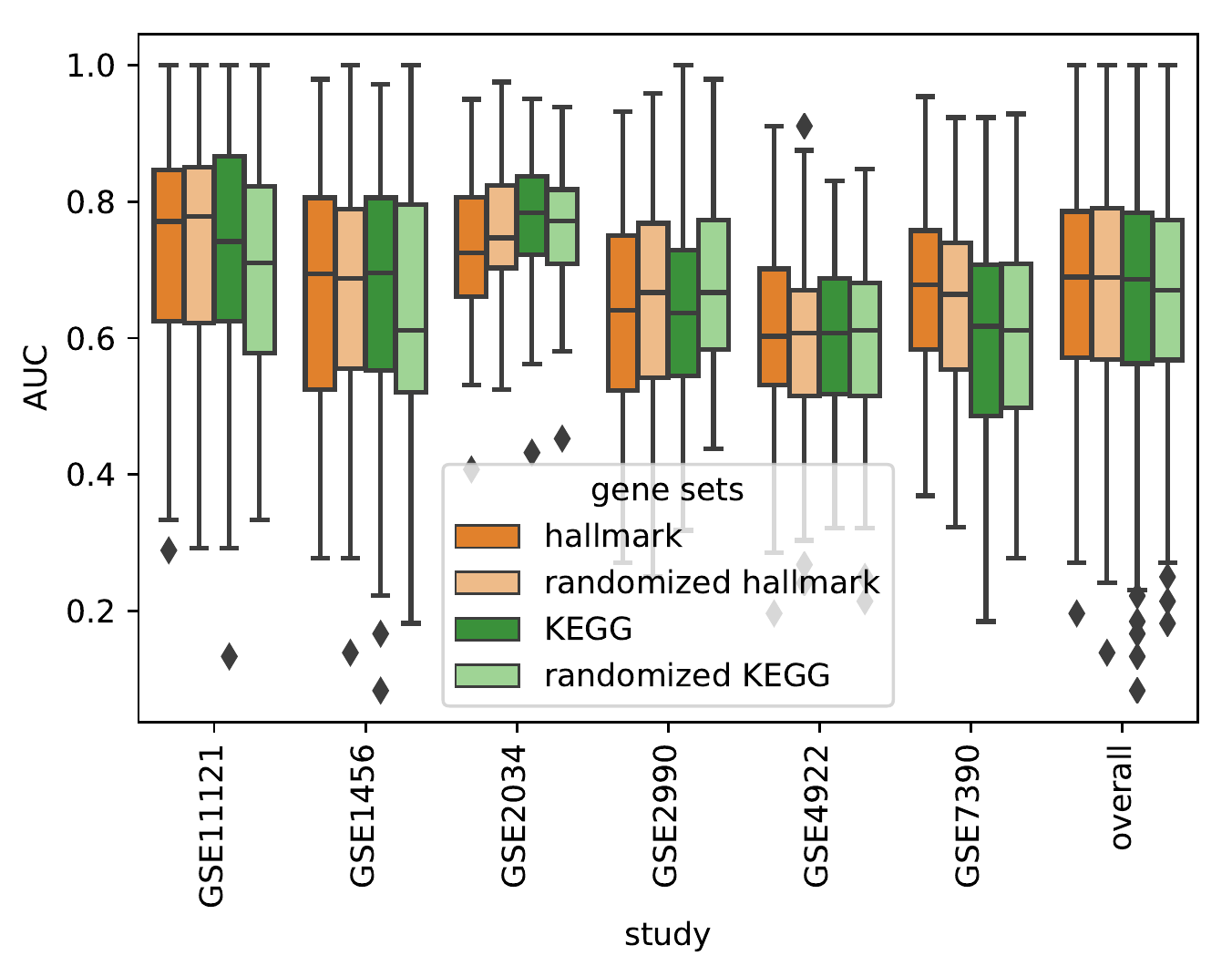}\\
\caption{
\label{sup:pimkl_cunetal_cv_genesets}
\textbf{ PIMKL cross-validation AUC for different gene sets.}
Box plots of all 100 AUC values (overall 600) for pathway induced MKL obtained by algorithm \ref{sup:cunetal_cv} with different gene sets to define the pathways given the same aforementioned interactions. In addition to the 50 previously introduced hallmark gene sets, results for 186 KEGG gene sets from the Molecular Signatures Database (MSigDB) version 5.2 \cite{Liberzon2015} and also respective randomized gene sets are reported. For randomization, the same number of gene sets is created, each set with random size between 50 and 250 genes by sampling from the union of all gene sets. The quartiles are comparable within each cohort proving the stability of the methods towards gene sets selection.
} 
\end{figure}

\begin{figure}[!htb]
\centering
\includegraphics[width=0.7\textwidth, clip]{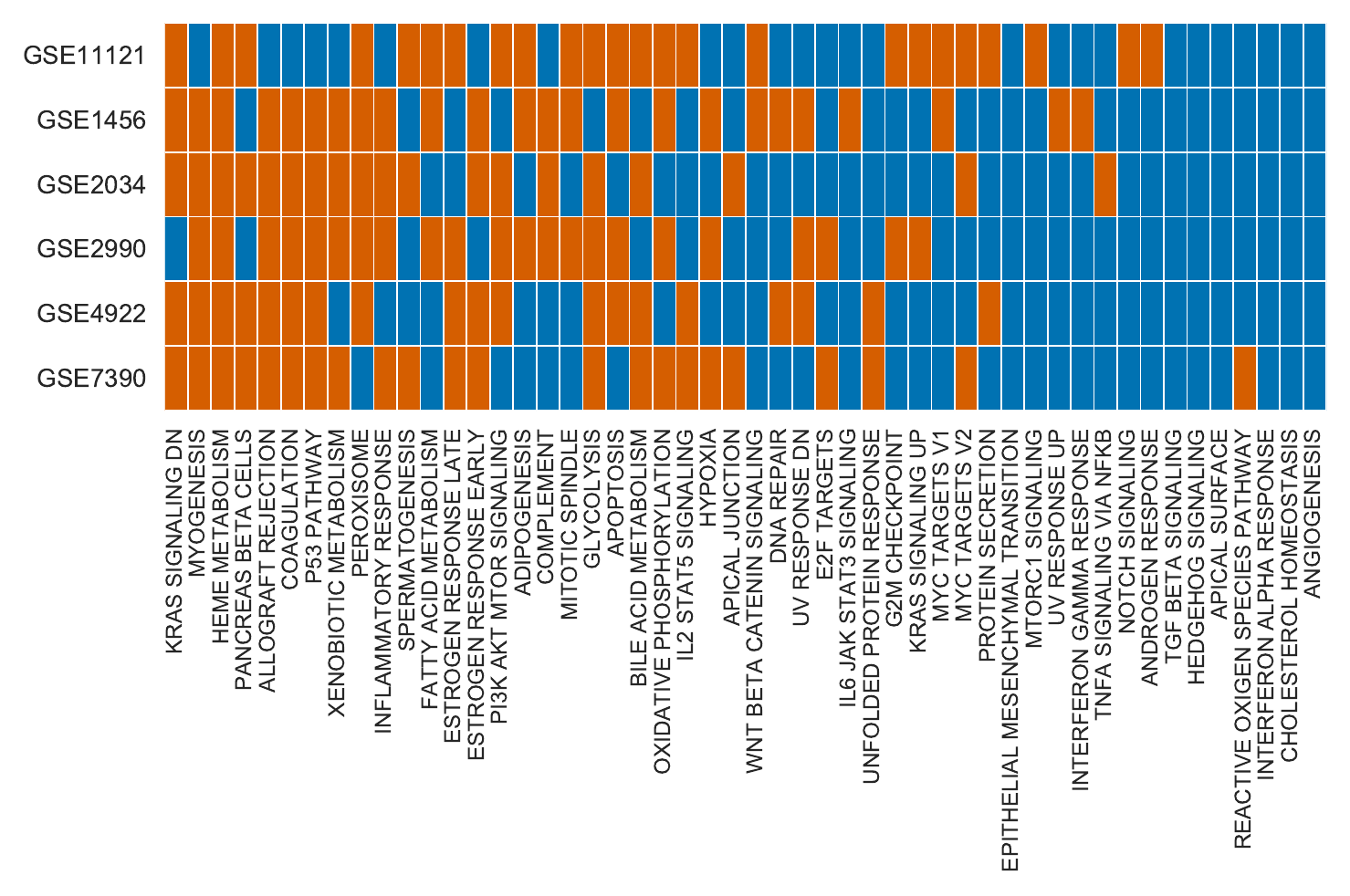}\\
\caption{
\label{sup:pimkl_cunetal_cv_cohorts_weights}
\textbf{ PIMKL cross-validation weights.}
Significance of weights over 100 cross-validation folds for the 50 hallmark pathways are reported. Significant pathways are colored in red, while non-significant in blue.
} 
\end{figure}

\end{document}